\documentclass[a4paper,11pt]{scrartcl}
\usepackage[centertags]{amsmath}
\usepackage{amsfonts}
\usepackage{amssymb}
\usepackage{amsthm}
\usepackage{nicefrac}
\usepackage{titlesec}
\usepackage[ruled, section]{algorithm}
\usepackage{pifont}
\usepackage[ansinew]{inputenc}
\usepackage{graphicx}
\usepackage{natbib}

\theoremstyle{plain}

\begin{document}

\title{A new stochastic differential equation modelling incidence and prevalence with an application to
systemic lupus erythematosus in England and Wales, 1995}

\author{Ralph Brinks\footnote{rbrinks@ddz.uni-duesseldorf.de}\\
Institute for Biometry and Epidemiology\\German Diabetes Center\\
Düsseldorf, Germany}

\date{}

\maketitle

\begin{abstract}
This article reformulates a common illness-death model in terms of a new
system of stochastical differential equations (SDEs). The SDEs are used to estimate epidemiological 
characteristics and burden of systemic lupus erythematosus in England and Wales in 1995.
\end{abstract}


\section{Introduction}
\begin{figure*}[ht]
\centerline{\includegraphics[keepaspectratio,
width=14cm]{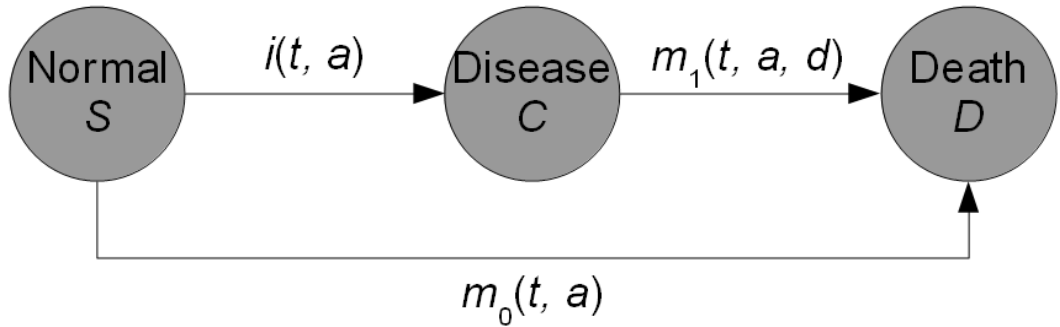}} \caption{Illness-death model 
of a chronic disease with three states. Persons in the state
\emph{Normal} are healthy with respect to the considered disease.
In the state \emph{Disease} they suffer from the disease. In the most
general case, the
transition rates depend on the calendar time $t,$ age $a,$ and
in case of the disease-specific mortality $m_1$ also on the
disease's duration $d$.} \label{fig:3states}
\end{figure*}

With a view to basic epidemiological parameters such as incidence, prevalence and mortality of a disease, 
it has proven useful to consider so called state models or compartmental models.
The model used here is also termed illness-death model \citep[Fig. 8.4]{Kal02}. It consists of
the three states \emph{Normal}, \emph{Disease}, \emph{Death} 
and the transitions between the states. \emph{Normal} means non-diseased with respect to the disease
under consideration. The numbers of persons in the \emph{Normal} and 
\emph{Disease} state are denoted as $S$ (susceptibles) and $C$ (cases), respectively.
The transition intensities (synonymously: rates) are called as shown in Figure \ref{fig:3states}: 
$i$ is the incidence rate, $m_0$ and $m_1$ are the mortality rates of the non-diseased 
and diseased persons, respectively. In general, the intensities depend on calendar time 
$t$, age $a$ and sometimes also on the duration $d$ of the disease.

When the rates do not depend on calendar time $t$, the model is called \emph{time-homogeneous}.
Then, with the additional condition that there is no dependency on the duration, 
\citeauthor{Mur94} have considered 
a two-dimensional system of ordinary differential equations (ODEs) to relate the changes
of the numbers of healthy and diseased persons with the rates of the in- and outflows of the 
corresponding states \citeyearpar{Mur94}\footnote{\citeauthor{Mur94} did not report the exact equations, 
but from a publication two years
later it may be deduced that they use an approach similar to Eq. \eqref{eq:MurrayODE}.}:
\begin{equation}\label{eq:MurrayODE}
\begin{split}
    \frac{\mathrm{d} S}{\mathrm{d} a} &= - \bigl ( i(a) + m_0(a) \bigr ) \cdot S\\
    \frac{\mathrm{d} C}{\mathrm{d} a} &= i(a) \cdot S - m_1(a) \cdot C.\\
\end{split}
\end{equation}

Age $a$ plays the role of temporal progression. The linear system \eqref{eq:MurrayODE} looks relatively 
harmless, but the impression is misleading.
Mostly only the age-specific mortality of the general population is well known,  
and rate $m_1$ is epidemiologically accessible as relative risk. Then, the system becomes nonlinear.

Furthermore, the inclusion of the hypothetical values $S$ and $C$ is disturbing. 
It would be better if we had the age-specific prevalence $p(a) := \tfrac{C(a)}{S(a) + C(a)}$ here, 
what indeed can be achieved \citep{Bri11}.

\bigskip

What are the benefits of such ODEs? For smooth incidence- and mortality rates 
plus an initial condition, the age profile of the numbers of patients or the prevalence 
is uniquely determined. To state it clearly, the ``forces'' incidence 
and mortality uniquely prescribe the prevalence – not only qualitatively but in these quantitative
terms. In this, we speak of the forward problem: we close from the causes – the forces 
– to the effect, namely the number of diseased persons. The reverse way, closing from the numbers 
of diseased persons
to the incidence, is the inverse problem -- we infer from the effect
to the cause. 

\bigskip

This paper is structured as follows: in the next section we describe the illness-death model of 
Figure \ref{fig:3states} in terms a new system of two stochastic differential equations (SDEs). 
As an application, in Section 3 we solve a forward problem to estimate the age-specific prevalence
of systemic lupus erythematosus (SLE) in England and Wales from published data. This allows
calculation of the mean age at onset of SLE, the mean duration and the burden of SLE in terms of 
diseased persons.

\section{Stochastic description of the illness-death model}
What can be achieved in the domain of ODEs, dividing the number $C(a)$ of the diseased by the number
$S(a) + C(a)$ of the living for deriving 
the prevalence at age $a$, is not that easy in random variables. Distributions of quotients 
of stochastically dependent random 
variables are problematic, so we have to model $S$ and $C$ bivariately. 

Let $X(a) := \left( S(a), C(a) \right )^t$ be the composite vector 
(the superscript denotes transposition).
For $\Delta a > 0$ define the vector $\Delta X$ of increments: 

$$\Delta X(a) := \bigl( S(a + \Delta a) - S(a), C(a + \Delta a) - C(a) \bigr )^t$$

Now we follow the reasoning of \citep{All99} and \citep{All08}, who have applied 
the theory presented here in the field of infectious diseases modeling. 

Choose $\Delta a > 0$ small that at most one person can change the state. 
In accordance with the definition of the rates $i$, $m_0$ and $m_1$ the following assumptions 
about the probability distribution $P(\Delta X(a))$ are made:

\begin{equation}\label{eq:probDist}
P \left ( \Delta X(a) =  {u \choose v} \right ) = 
  \begin{cases}
    m_0(a) \cdot S(a) \cdot \Delta a + o( \Delta a ) & \textnormal{if } (u, v) = (-1, 0)\\
    m_1(a) \cdot C(a) \cdot \Delta a + o( \Delta a ) & \textnormal{if } (u, v) = (0, -1)\\
    i(a)   \cdot S(a) \cdot \Delta a + o( \Delta a ) & \textnormal{if } (u, v) = (-1, 1)\\
    1 - \bigl [ m_0(a) \cdot S(a) + m_1(a) \cdot C(a) +  & \\
    ~~~~~~~ i(a) \cdot S(a) \bigr ] \cdot \Delta a + o( \Delta a ) & \textnormal{if } (u, v) = (0, 0)\\
  \end{cases}
\end{equation}

If we further assume that the increments are normally distributed, 
we get the expected value 
\begin{equation*}
E \bigl ( \Delta X(a) \bigr ) =  \left [  m_0(a) \cdot S(a) \cdot {-1 \choose 0} 
                        + m_1(a) \cdot C(a) \cdot {0 \choose -1}
                        + i(a)   \cdot S(a) \cdot {-1 \choose 1}  \right ] \cdot \Delta a + o(\Delta a)
\end{equation*}
and covariance matrix
\begin{equation*}
\begin{split}
V \bigl ( \Delta X(a) \bigr ) &= E \left (  \Delta X(a) \cdot \Delta X^t(a) \right ) - 
                                E \left (  \Delta X(a) \right ) \cdot E \left ( \Delta X^t(a) \right ) \\
                               &\approx E \left (  \Delta X(a) \cdot \Delta X^t(a) \right ).
\end{split}
\end{equation*}
The matrix $V$ is symmetric and positively definite. Hence, there is uniquely determined 
matrix square root $V^{1/2}$. Due the normal distribution assumption the vector $X$ 
fulfills

\begin{equation}\label{eq:Approx}
X(a + \Delta a) = X(a) + \Delta X(a) = X(a) + E \bigl ( \Delta X(a) \bigr ) + V \bigl ( \Delta X(a) \bigr )^{1/2} \, \xi, 
\end{equation}
where $\xi = (\xi_1, \xi_2)^t$ has normally distributed components $\xi_i \sim N(0, 1), ~i=1, 2.$

\bigskip

Under certain smoothness conditions about the coefficient functions $i, m_1$ and $m_0$ 
the difference equation \eqref{eq:Approx} is an Euler approximation to the Îto SDE system 

\begin{equation}\label{eq:SDE}
\begin{split}
    \frac{\mathrm{d} S}{\mathrm{d} a} &= - \bigl ( i(a) + m_0(a) \bigl ) \cdot S 
                                       + b_{11} \frac{\mathrm{d} W_1}{\mathrm{d} a}
                                       + b_{12} \frac{\mathrm{d} W_2}{\mathrm{d} a}\\
    \frac{\mathrm{d} C}{\mathrm{d} a} &= i(a) \cdot S - m_1(a) \cdot C
                                       + b_{21} \frac{\mathrm{d} W_1}{\mathrm{d} a}
                                       + b_{22} \frac{\mathrm{d} W_2}{\mathrm{d} a}.\\
\end{split}
\end{equation}

\noindent In this expression, $W_1$ and $W_2$ are independent Wiener processes \citep{Klo99}
and the 
$2 \times 2$ -- matrix $B = (b_{ij})$ is the
uniquely determined square root of the covariance matrix divided by $\Delta a$: 
$$B = \bigl ( V(\Delta X) / \Delta a \bigr )^{1/2}.$$

\bigskip
 
Which advantages has the SDE formulation compared to the ODE? 
In rare diseases as in the next section, the inclusion 
of uncertainty is sometimes more appropriate than calculating deterministically. 
In addition, SDEs sometimes have properties that cannot 
be derived from the theory of ODEs, as for example the quasistationary solutions \citep{Dar67}.

\section{Application to Systemic Lupus Erythematosus}
In this section the SDE is applied to epidemiological data of systemic lupus erythematosus
in England and Wales. Systemic lupus erythematosus (SLE) is a severe rheumatic disease
with a variety of clinical manifestations. Despite several therapy options, patients often 
are restricted heavily in quality of life and ability to work. Epidemiological data are rare. 
Here, the incidence data for males and females is taken from the UK General Practice 
Research Database (GPRD) in the years 1990--1999 as reported in \citep{Som07}. Mortality $m_1$ of 
SLE patients
is modeled by the relative mortality as reported in \citep{Ber06}. Duration of SLE was not taken
into account.

Regarding the mortality of the non-diseased, we take the mortality in the 
general population. Due to the low prevalence of 
SLE this is legitimate. Then, 5000 solution paths of the SDE system \eqref{eq:SDE} 
are simulated by the Euler-Maruyama method \citep{Klo99} and the corresponding 
age-specific prevalences have been calculated. This is done for males and females separately.

\begin{figure*}[ht]
\centerline{\includegraphics[keepaspectratio,
width=.6\linewidth]{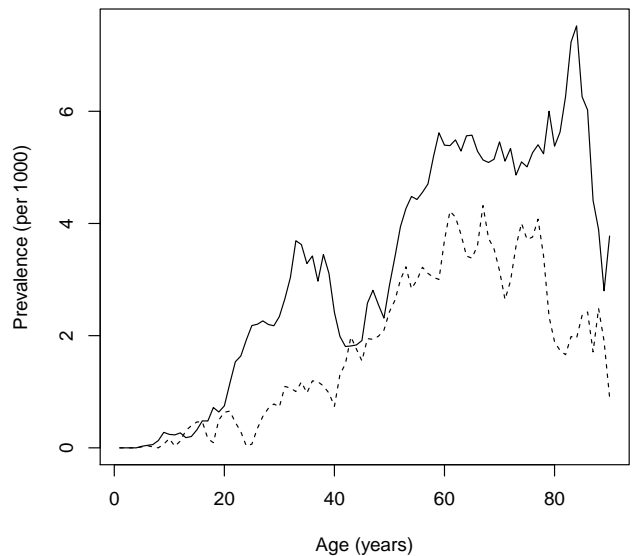}} \caption{Example paths of the age-specific 
prevalence of SLE resulting from two pairs of solution paths of system \eqref{eq:SDE}.} 
\label{fig:TwoPaths}
\end{figure*}

As an example, Figure \ref{fig:TwoPaths} shows the prevalence resulting from two pairs 
of solution paths.

\clearpage

\begin{figure*}[ht]
  \begin{minipage}[b]{.48\linewidth}
    \includegraphics[width=\linewidth]{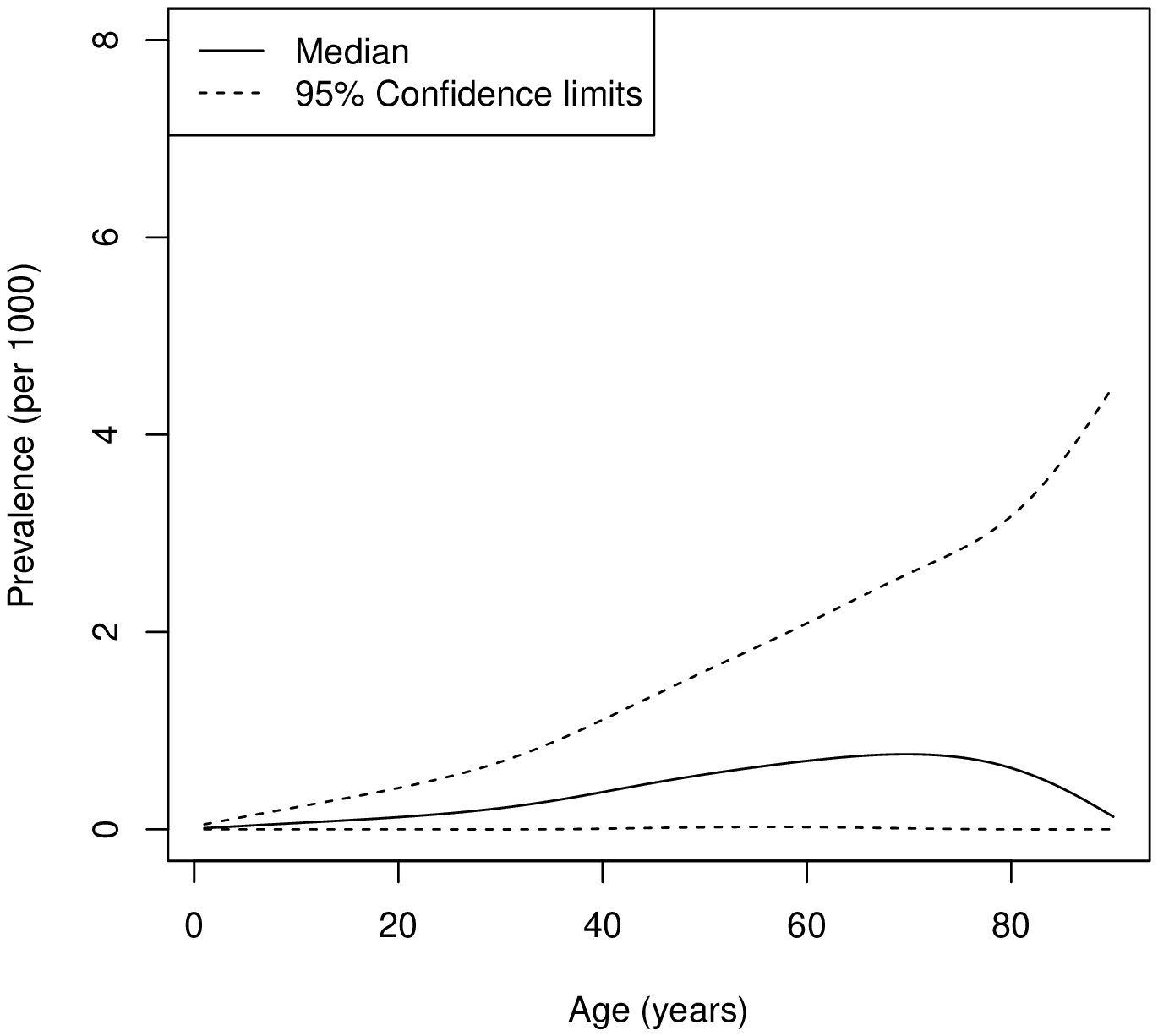}
    \caption{Age-specific prevalence of SLE in males.}\label{fig:PrevMal}
  \end{minipage}
  \hspace{.04\linewidth}
  \begin{minipage}[b]{.48\linewidth}
    \includegraphics[width=\linewidth]{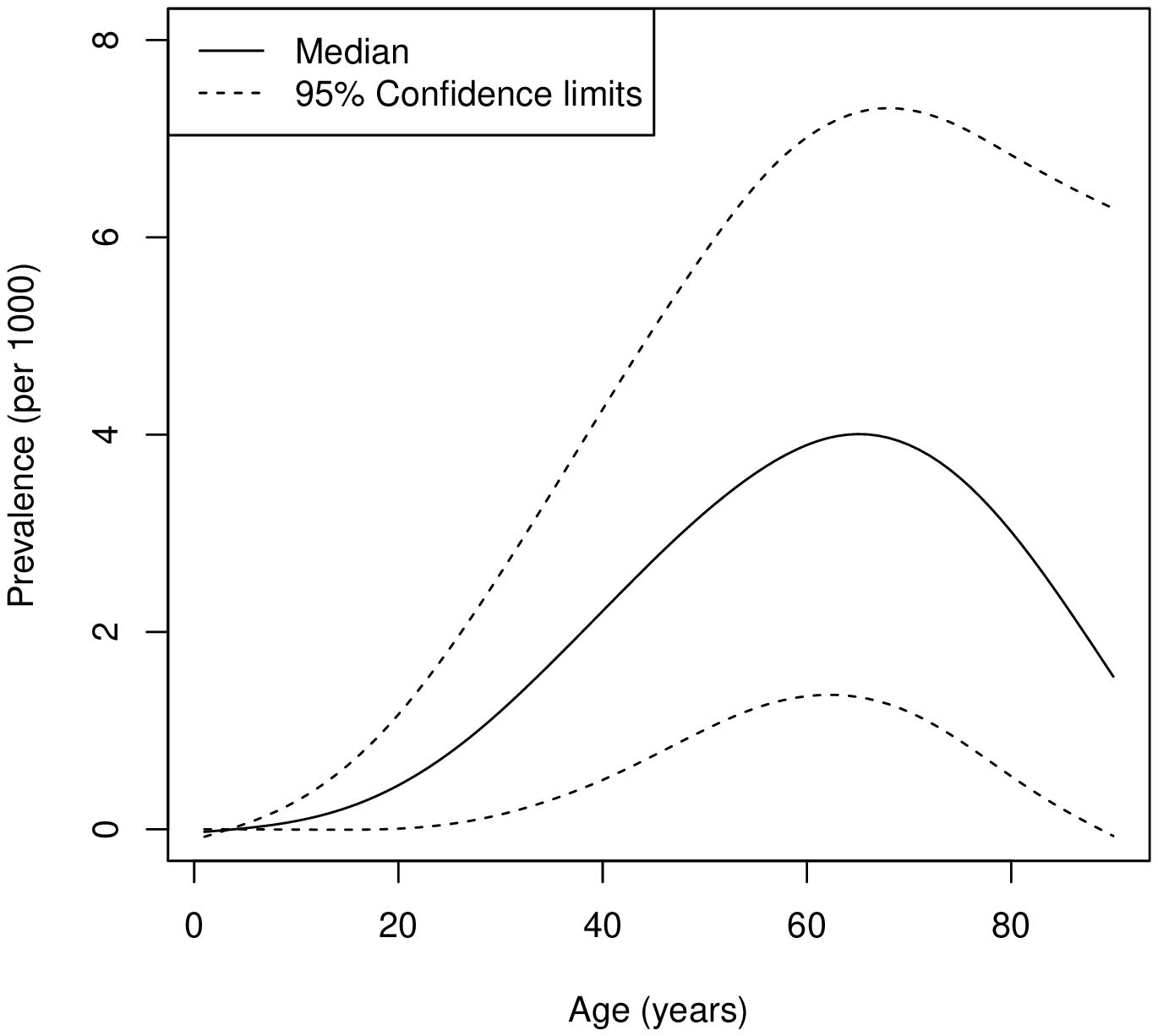}
    \caption{Age-specific prevalence of SLE in females.}\label{fig:PrevFem}
  \end{minipage}
\end{figure*}

Of course, single paths for the prevalence are not that important. It is more interesting, 
to analyze where the paths of the prevalence lie and what the charcteristics are. As an
example, Figures \ref{fig:PrevMal} and \ref{fig:PrevFem} show the regions where 95\% of the
5000 solution paths lie. The upper and the lower dotted curve indicate the 97.5\% and 2.5\%
quantile of the 5000 prevalence paths, respectively. This means, for each age $a$ the corresponding
quantiles from the empirical distribition of the 5000 values at age $a$ are calculated. 
Additionally, Figures \ref{fig:PrevMal} and \ref{fig:PrevFem} show the curve of the median 
(solid line).

The median curves of males and females indicate the big difference of prevalent SLE between males
and females. This is due to the fact, that gender is a risk factor for SLE and incidence between
males and females differ strongly. The hazard ratio (females vs. males) is about 10 in the age-group 
of 25-35 years. The hazard ratio
decreases to about 5 in the following age-classes until 65 years and after that lowers to about 2.

For an estimate of the burden of SLE in England and Wales one may estimate the total number $\mathfrak{C}$ of persons with
SLE:

\begin{equation}\label{eq:OverallC}
\mathfrak{C} = \sum_{a=0}^{90} p(a) \cdot N(a),
\end{equation}
where $N(a)$ denotes the number of persons in England and Wales aged $a = 0, \dots, 90.$ The number $N(a)$ is 
obtained from official
vital statistics in the year 1995, \citep{ONS11}. The age-specific prevalence $p(a)$ is taken from the 5000 paths.

\begin{figure*}[ht]
  \begin{minipage}[b]{.48\linewidth}
    \includegraphics[width=\linewidth]{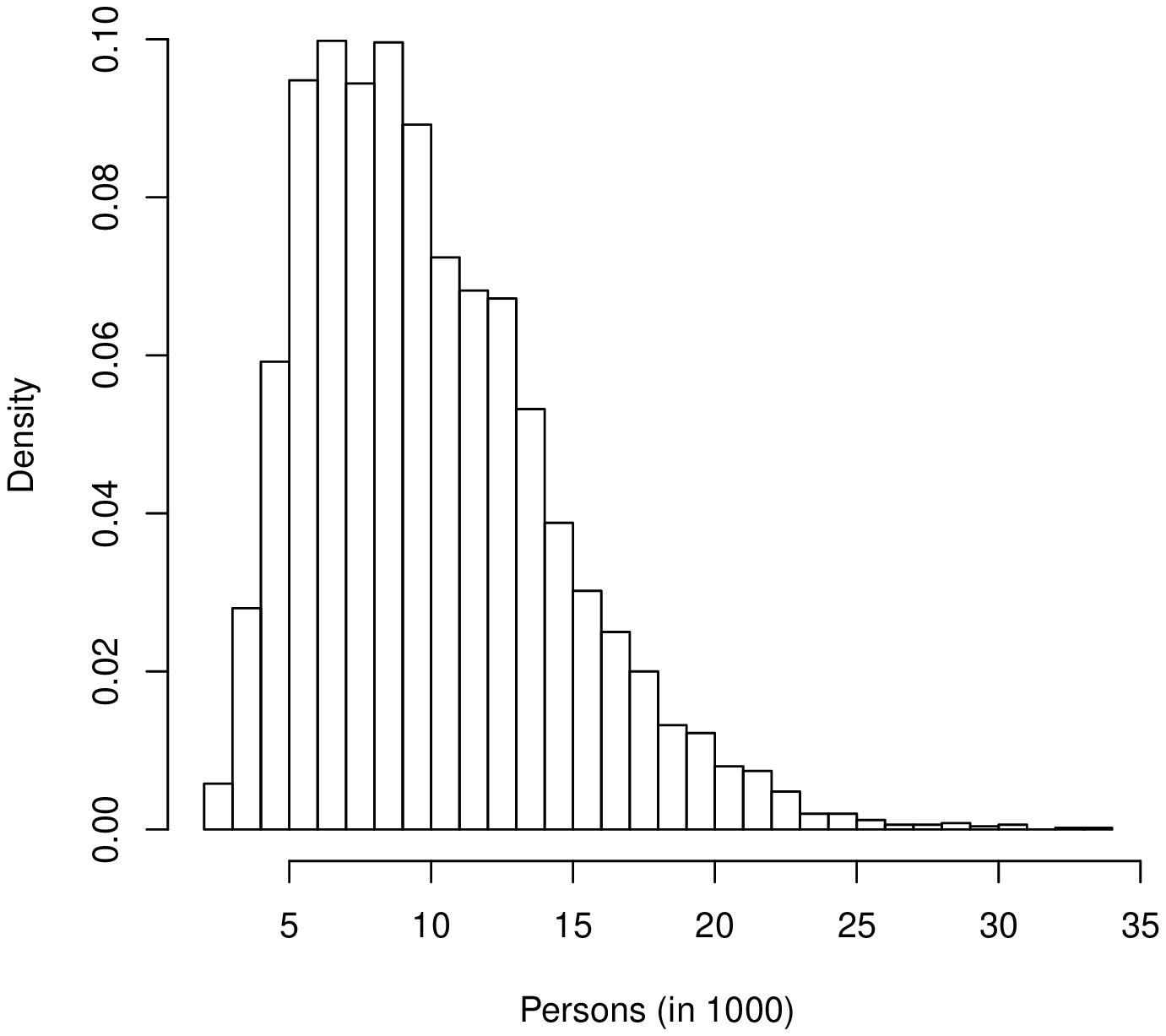}
    \caption{Histogram of the number of males with SLE.}\label{fig:C_m}
  \end{minipage}
  \hspace{.04\linewidth}
  \begin{minipage}[b]{.48\linewidth}
    \includegraphics[width=\linewidth]{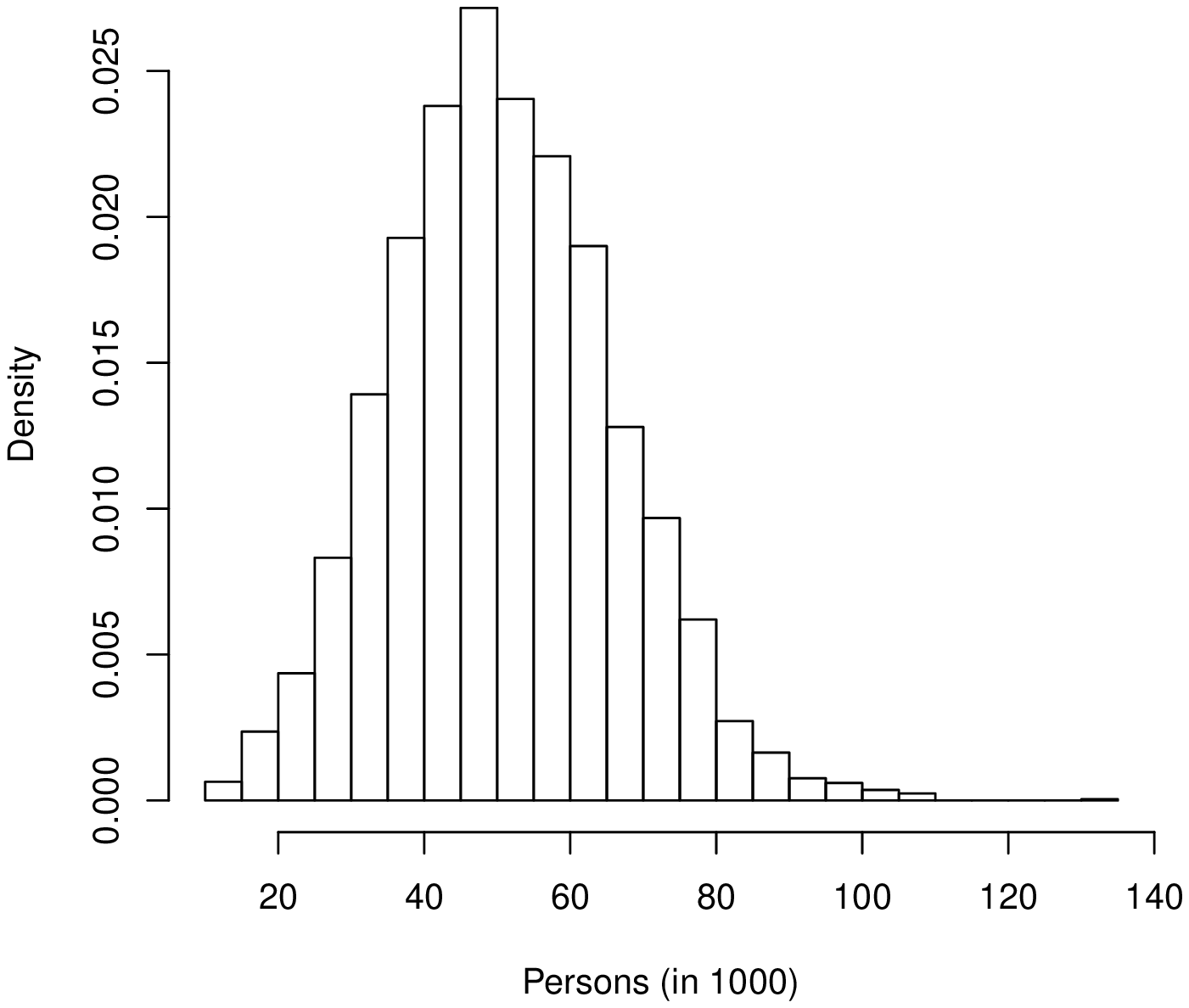}
    \caption{Histogram of the number of females with SLE.}\label{fig:C_f}
  \end{minipage}
\end{figure*}

Figures \ref{fig:C_m} and \ref{fig:C_f} show the distributions of $\mathfrak{C}$ obtained from the 5000 paths.
Again, the enormous difference between males and females becomes obvious. While 50.1 thousand 
females are affected in England and Wales (interquartile range (IQR): 40.3--60.9), for males
the corresponding number is 9.2 (IQR: 6.6--12.5) thousand.

In situation described here, it is possible to derive the mean duration $\mathfrak{D}$ of SLE in males and females.
The mean duration $\mathfrak{D}$ is the number of person-years of all SLE patients divided by the total number of 
persons who ever got it:
\begin{equation}\label{eq:MeanDuration}
\mathfrak{D} = \frac{\sum\limits_{a=0}^{90} p(a) \cdot N(a)}{\sum\limits_{a=0}^{90} i(a) \cdot \bigl ( 1-p(a) \bigr ) \cdot N(a)}.
\end{equation}
If we calculate this value for all paths $p$, we find that in males and females the mean duration is
23.2 (IQR: 16.7--31.5) and 23.9 (IQR: 16.2--29.1) years, respectively. Thus, genders do not differ much in 
that respect. Similarly, the mean age at onset $\mathfrak{M}$ may be computed:
\begin{equation}\label{eq:MAO}
\mathfrak{M} = \frac{\sum\limits_{a=0}^{90} a \cdot i(a) \cdot \bigl ( 1-p(a) \bigr ) \cdot N(a)}
{\sum\limits_{a=0}^{90} i(a) \cdot \bigl ( 1-p(a) \bigr ) \cdot N(a)}.
\end{equation}

The empirical distribution of $\mathfrak{M}$ in the 5000 paths yields 51.750 (IQR: 51.748--51.752) and 46.108 
(46.102--46.113) years in males and females, respectively. It is striking that $\mathfrak{M}$ has a relativly 
low variability in both genders. This is due to the factor $1-p(a)$, which is close to unity.


\section{Conclusion}
In the domain of infectious diseases, the theory of deterministic 
differential equations was generalized towards stochastic differential equations
more than a decade ago, \citep{All99}. In this article this 
transformation has been accomplished in the field of chronic diseases. The numbers of 
healthy and diseased persons have been modeled by a new system of two Îto stochastic
differential equations. 
In rare chronic diseases such as systemic lupus erythematosus, a stochastic 
formulation might be preferable over a deterministic. Even if the incidence and mortality
rates are well-known, statistical fluctuations in the number of diseased have a strong
impact in the age course of the prevalence. This becomes obvious in Figures \ref{fig:C_m} and
\ref{fig:C_f} where the distribution of the total number of males and females with SLE 
in England and Wales in 1995 have been estimated. The middle fifty spans about 6 and 20 thousand
males and females respectively. Additionally, other disease characteristics have been calculated. 
The mean age at onset as derived in our theoretical model is 51.8 and 46.1 for males and females,
respectively. The corrsponding empirical values 52.2 and 46.3 observed in the register data by 
\citeauthor{Som07} are in good agreement. Another hint for the appropriateness of the methods described here
comes from the basic epidemiological equation, that overall prevalence equals the product of overall incidence
and duration of the disease \citep{Szk07}. The overall prevalence in males and females can
easily be obtained by Eq. \eqref{eq:OverallC} and the age pyramid $N(a)$. In our model the overall
prevalence divided by the mean duration (Eq. \eqref{eq:MeanDuration}) yields the overall incidence 
$1.58$ and $7.95$ per 100000 person-years for males and females, respectively. Again, this is close
to the empirically observed values $1.60$ and $8.01$ per 100000 person-years \citep[Tab. 1]{Som07}.
When relating the results of this study to other epidemiological data from the UK, especially the overall
incidence for females appears high. \citeauthor{Hop93} find a value of about $6.5$ per 100000 person-years 
only \citeyearpar{Hop93}. However, it has to be noted that the data of \citep{Hop93} are in a way inconsistent:
If we calculate the mean duration of SLE in females by the basic epidemiological equation for the data
of \citeauthor{Hop93},
we find a mean duration of about 7 years only, which contradicts common survival times of persons 
with SLE \citep{Cer03}.

\bigskip

Although there is an ongoing debate about the differences between childhood-onset SLE and adult-onset
as well as duration of SLE as a risk factor for comorbidities and mortality, it has to be noticed 
that reliable epidemiologic data 
about age at onset, duration of the disease, mean age of diseased etc are sparse or lacking. As an example,
the sytematic review \citep{Dan06} about the global burden and epidemiology of SLE just found one study from 
Germany, the European country with the most inhabitants. The associated publication \citep{Zin01} was about 
prevalent cases, but did not mention a prevalence estimate. Incidence had not been adressed.

Theoretical models such as the one presented here may help to at least roughly 
estimate the burden and characteristics of rare chronic diseases. This is especially true in countries
with few epidemiological or administrative data. However, the approach described here has several limitations. 
First, the stochastic differential equation 
\eqref{eq:SDE} does not take into account calendar time trends. In the application to SLE, it has been
shown that relative mortality of SLE patients undergoes a secular trend, \citep[Tab. 6]{Ber06}. In the same
publication we find, that relative mortality in persons with SLE depends on the duration of the 
disease. The longer a person is diseased, the more the relative risk decreases. Duration dependence 
is not modeled in Eq. \eqref{eq:SDE}. Hence, the new approach may be used as an approximation only, 
and more evaluations of the method are necessary to examine validity and applicability of the model. However, the
disease characteristics derived by the new methods in this work are consistent and indicate an interesting 
and maybe fruitful way to go.

\end{document}